\documentclass[aps,prd,preprint,superscriptaddress]{revtex4}
\usepackage{amsmath}
\usepackage{amssymb}
\usepackage{epsfig}
\usepackage{ulem}
\usepackage{graphics}
\usepackage{indentfirst}
\usepackage{color}
\usepackage{multirow}

\newcommand{\pb}{\ensuremath{\bar{\Phi}}}

\newcommand{\te}{\ensuremath{\theta}}
\newcounter{RomanNumber}

\newcommand{\rr}{\ensuremath{\textbf{r}}}

\begin{document}

\preprint{ACT-01-15, MI-TH-1506}

\vspace*{-0.6 cm}
\title{Symmetry Breaking Indication for Supergravity Inflation in Light of the Planck $2015$\vspace*{0.2cm}}

\author{Tianjun Li \vspace*{0.05 cm}}
\affiliation{{\footnotesize State Key Laboratory of Theoretical Physics
and Kavli Institute for Theoretical Physics China (KITPC),
      Institute of Theoretical Physics, Chinese Academy of Sciences,
Beijing 100190, P. R. China}}

\affiliation{{\footnotesize School of Physical Electronics,
University of Electronic Science and Technology of China,
Chengdu 610054, P. R. China}}

\author{Zhijin Li}

\affiliation{{\footnotesize George P. and Cynthia W. Mitchell Institute for
Fundamental Physics and Astronomy,
Texas A\&M University, College Station, TX 77843, USA}}

\author{Dimitri V. Nanopoulos}

\affiliation{{\footnotesize George P. and Cynthia W. Mitchell Institute for
Fundamental Physics and Astronomy,
Texas A\&M University, College Station, TX 77843, USA}}

\affiliation{{\footnotesize Astroparticle Physics Group, Houston Advanced
Research Center (HARC), Mitchell Campus, Woodlands, TX 77381, USA}}

\affiliation{{\footnotesize Academy of Athens, Division of Natural Sciences,
28 Panepistimiou Avenue, Athens 10679, Greece} \vspace*{0.1cm}}

\begin{abstract}

The supergravity (SUGRA) theories with exact global $U(1)$ symmetry or shift symmetry in K\"ahler potential provide
the natural frameworks for inflation. However, the quadratic inflation is disfavoured by the new results
on primordial tensor fluctuations from the Planck Collaboration. To be consistent with the new Planck data,
we point out that the explicit symmetry breaking is needed, and study these two SUGRA inflation
in details. For the SUGRA inflation with global $U(1)$ symmetry, the symmetry breaking
term leads to a trigonometric modulation on inflaton potential. The coefficient of the $U(1)$ symmetry breaking term
is of the order $10^{-2}$, which is sufficient large to improve the inflationary predictions while its higher order
corrections are negligible. Such models predict sizeable tensor fluctuations and highly agree with the Planck results.
In particular, the model with a linear $U(1)$ symmetry breaking term predicts the tensor-to-scalar ratio around
$\rr\sim0.01$ and running spectral index $\alpha_s\sim-0.004$, which comfortably fit with the Planck observations.
For the SUGRA inflation with breaking shift symmetry, the inflaton potential is modulated by an exponential
factor. The modulated linear and quadratic models are consistent with the Planck observations. In both kinds of models
the tensor-to-scalar ratio can be of the order $10^{-2}$, which will be tested by the near future observations.

\end{abstract}

%\pacs{04.65.+e, 04.50.Kd, 12.60.Jv, 98.80.Cq}

\maketitle

\section{Introduction}

To realize inflation \cite{Guth:1980zm, Linde:1982} in supergravity (SUGRA) theory, the flat conditions give
strong constraints on the F-term scalar potential with an exponential factor $e^{K(\Phi,\pb)}$ which is too steep
to generate inflation by the field close or above the reduced Planck scale. This is the well-known $\eta$ problem
for SUGRA inflation. The $\eta$ problem can be solved if the K\"ahler potential admits certain symmetry, such as in no-scale SUGRA
with global $SU(N,1)/SU(N)\times U(1)$ symmetry \cite{Cremmer:1983bf}.
The classical quadratic inflation \cite{Linde:1982} is simply realized in supergravity theory
 with global $U(1)$ symmetry \cite{Li:2014vpa, Li:2014unh} or shift symmetry \cite{Kawasaki:2000yn}
in the K\"ahler potential. The quadratic inflation predicts large primordial tensor fluctuations with
the tensor-to-scalar ratio $\rr\simeq0.15$. However, the recent observations from the
Planck and BICEP2/Keck Array Collaborations have provided strong constraints on the primordial tensor
fluctuations~\cite{Planck:2015xua, Ade:2015oja, Ade:2015tva}, $\rr<0.11$ ($\rr<0.12$ from BICEP2/Keck Array)
at $95\%$ Confidence Level (C.L.). So the simplest quadratic infaltion is disfavoured. In light of
the new Planck results, another proposal for supergravity inflation with slightly explicit symmetry breaking
becomes important. This is based on the fact that the $\eta$ problem can also be solved by an approximate symmetry
in the K\"ahler potential while the inflationary observables are sensitive to such potential variation.
Interestingly, the inflationary observables can be significantly modified while the models are still free from
the $\eta$ problem.

A natural solution to the $\eta$ problem is from the global $U(1)$ symmetry in the K\"ahler potential of
the minimal supergravity (mSUGRA) $K=\Phi\pb$, which is invariant under the $U(1)$ rotation:
$\Phi\rightarrow\Phi e^{i\te}$. To employ the phase $\theta$ as the inflaton, it requires strong field stabilization
in the radial direction and phase monodromy in the superpotential, which are simply fulfilled in helical phase
inflation driven by the potential with helicoid structure \cite{Li:2014vpa}.
The global $U(1)$ symmetry is of specially importance,  because it not only provides a new solution for
the $\eta$ problem, but also protects the models away from quantum loop corrections, which can appear only
in the K\"ahler potential but not in the superpotential, and they depend on
the radial component instead of the phase so have little effect on the phase inflation. Moreover,
according to the Lyth bound \cite{Lyth:1996im} on the inflationary models with
tensor-to-scalar ratio larger than 0.01, the super-Planckian field excursion is needed, which potentially
makes the models unreliable due to the quantum gravity effects.
In the helical phase inflation, the super-Planckian field excursion is fulfilled
along a helix trajectory, and then the problem is solved.
Moreover, the helical phase inflation can realize a super-Planckian phase decay constant through
 phase monodromy, which corresponds to the explicit $U(1)$ symmetry breaking in the superpotential and provides a simple phase-axion alignment.

The helical phase inflation remarkably relates to many important and interesting developments on large field inflation.
The idea to employ the phase, a pseudo-Nambu-Goldstone boson (PNGB) as inflaton was first proposed
in Ref.~\cite{Freese:1990rb} to protect the flat potential away from the Ultra Violet (UV) corrections.
The inflation driven by the PNGB potential has been studied extensively in
Refs.~\cite{German:2001sm, Baumann:2010nu, Harigaya:2014eta, McDonald:2014oza, Carone:2014cta, Barenboim:2014vea, McDonald:2014rha, Barenboim:2015zka}. The axion alignment mechanism for super-Planckian axion decay constant was first proposed in
Ref.~\cite{Kim:2004rp}. The monodromy inflation as an attractive proposal to realize the super-Planckian
field excursion in string theory was provided in Refs.~\cite{Silverstein:2008sg, Kaloper:2008fb}.
In Refs.~\cite{Choi:2014rja, Tye:2014tja, Kappl:2014lra} the axion alignment mechanism is explained
as a special type of monodromy inflation realized by axions. In fact, a similar name ``helical inflation" was first
used in Ref.~\cite{Tye:2014tja}, in which the helical structure refers to the alignment of axions,
while in our models the ``helical" path is from the single phase component of a complex field with stabilized radial component.

Considering the crucial role of the global $U(1)$ symmetry for inflation, it is questionable if the merits of helical phase
inflation maintain after the $U(1)$ symmetry breaking. Since the global $U(1)$ symmetry is broken at tree level,
the extra terms may be generated from quantum loop corrections or after integrating out heavy fields and then
break the $U(1)$ symmetry further. To fit the experimental observations, the inflationary observables are of
the order $10^{-2}$. So the $U(1)$ symmetry breaking is at the same order, which is large enough to improve
the inflation predictions while its higher order corrections are too small to induce notable effect.
In this work we will show that by introducing a small $U(1)$ symmetry breaking term, the phase potential
will be slightly modulated by a trigonometric factor, and their predictions on inflation are highly
consistent with the new Planck results. Specifically, it predicts
interesting running of spectral index with magnitudes depending on the $U(1)$ symmetry breaking term.

Another concise solution of the $\eta$ problem has been proposed for years \cite{Kawasaki:2000yn}.
Its K\"ahler potential is constrained by an extra shift symmetry: $\Phi\rightarrow\Phi+iC$
so that $K=K(\Phi+\pb)$ is independent of the Im$(\Phi)$ and then the $\eta$ problem is evaded for
the inflation driven by Im$(\Phi)$.  Interestingly, this kind of models was shown to be closely related to
the helical phase inflation models with $U(1)$ symmetry, as studied in \cite{Li:2014unh}.
Through a field redefinition $\Phi=e^{\Psi}$ and ignoring the higher order terms which vanish after
field stabilization, the models with $U(1)$ symmetry reduce to the models
 with shift symmetry $\Psi\rightarrow\Psi+iC$.

In Ref.~\cite{Li:2013nfa} we for the first time proposed the shift symmetry breaking in the K\"ahler potential.
The quadratic potential, as well as other power-law potentials are modulated by an exponential factor,
which generates inflation with a scalar spectral index $n_s\approx0.96\sim0.97$ and especially
a broad range of tensor-to-scalar ratio $\rr$. The predictions are well consistent with the Planck results
published in $2013$~\cite{Ade:2013uln}. The potential role of the symmetry breaking term was discussed
in Ref.~\cite{Kallosh:2010ug} and the tensor-to-scalar ratio can be as large as $\rr\simeq0.2$, as shown
in Ref.~\cite{Harigaya:2014qza} in light of the BICEP2 results on large tensor fluctuations evaluated
from B-mode polarizations that are significantly affected by the dust contributions~\cite{Ade:2014xna}.
Because the new Planck results provided stronger constraint on the tensor-to-scalar ratio,
it is very important to compare these models with the new observations further.

In this work, we study SUGRA inflation with breaking global $U(1)$ symmetry or shift symmetry
in the K\"ahler potential, and compare their predictions with the new Planck results.
 In the SUGRA inflation with global $U(1)$ symmetry, a trigonometric modulation on inflaton potential
is generated from the symmetry breaking term. The coefficient of the $U(1)$ symmetry breaking term
at order $10^{-2}$ is large enough to improve the inflationary predictions while the higher order
corrections are negligible.  The predicted sizeable tensor fluctuations are highly consistent
with the Planck results.  Especially, the model with a linear $U(1)$ symmetry breaking term has
the tensor-to-scalar ratio around
$\rr\sim0.01$ and running spectral index $\alpha_s\sim-0.004$, which comfortably fit with the Planck observations.
In the SUGRA inflation with breaking shift symmetry,  the modulated linear and quadratic models agree with
the Planck observations due to the additional exponential
factor in the inflaton potential. In these two kinds of models,
the tensor-to-scalar ratio can be of the order $10^{-2}$, which will be tested by the near future experiments.
Therefore, the new Planck data strongly suggest the global $U(1)$ or shift symmetry breaking in K\"ahler
potential for SUGRA inflation.

This paper is organized as follows. In Section II we review the helical phase inflation based on
the minimal supergravity with global $U(1)$ symmetry. In Sections III and IV, we study the inflationary models
associated with $U(1)$ symmetry breaking and shift symmetry breaking, respectively.
Our conclusion is given in Section V.

\section{Brief Review of Helical Phase Inflation}

Initially the helical phase inflation was introduced to solve the $\eta$ problem for inflation
in supergravity theory~\cite{Li:2014vpa, Li:2014unh}. It starts from the trivial fact that in the minimal supergravity,
the K\"ahler potential for a chiral superfield $\Phi$: $K(\Phi,\pb)=\Phi\pb$ is invariant under a global $U(1)$
transformation: $\Phi\rightarrow\Phi e^{i\theta}$, and so is the factor $e^K$ in the F-term scalar potential. In consequence,
by using the phase component of $\Phi$ as inflaton the $\eta$ problem is solved automatically. Nevertheless,
we need to solve two problems related to this proposal:  stabilization of the radial component of $\Phi$
and realization of the phase monodromy in superpotential.

Firstly, the radial component of $\Phi$ should be stabilized otherwise it would generate
notable iso-curvature perturbations that contradict with the experimental observations. However,
it is non-trivial to stabilize the norm of $\Phi$ while keep its phase light as they couple with each other.
Because the superpotential $W(\Phi)$ is an holomorphic function of $\Phi$, without extra $U(1)$ charged field
$W$ is not invariant under general $U(1)$ transormation. If, for a whole circular $U(1)$ rotation
$\Phi\rightarrow\Phi e^{i2\pi}$, the superpotential $W$ is invariant, then the F-term scalar potential of $\Phi$
is exactly periodic under $\theta\rightarrow\theta+2\pi$. With a sub-Planckian field norm $|\Phi|\leqslant M_P$,
such potential cannot provide sufficient trans-Planckian field excursion that is needed for large field inflation
with tensor-to-scalar ratio $\rr>0.01$. Therefore, the superpotential $W$ should break the global $U(1)$ symmetry
in the way without the discrete symmetry $\Phi\rightarrow\Phi e^{2\pi i}$. In the other words, there is
a phase monodromy in $W$. For different phase monodromies in $W$, we can get different types of large field inflationary models,
for example, the quadratic inflation or natural inflation with super-Planckian decay constant.

\subsection{The Quadratic Inflation}

The helical phase inflation with quadratic potential was proposed in Ref.~\cite{Li:2014unh}. The K\"ahler potential
and superpotential in supergravity theory are
\begin{equation}
K=\Phi\pb+X\bar{X}-g(X\bar{X})^2,~~~~  W=a\frac{X}{\Phi}\ln(\Phi). \label{sup1}
\end{equation}
The global $U(1)$ symmetry is broken by the superpotential while restored when $a=0$, so the global $U(1)$ symmetry is technically natural \cite{tHooft}. The phase monodromy in the superpotential $W$ is
\begin{equation}
\Phi\rightarrow \Phi e^{2\pi i}, ~~~~W\rightarrow W+2\pi ai\frac{X}{\Phi}. \label{phm1}
\end{equation}
As shown in Refs.~\cite{Li:2014vpa} and \cite{Li:2014unh}, the above
superpotential $W$ in Eq.~(\ref{sup1}) can be obtained from the following superpotential
\begin{equation}
W_0=\sigma X\Psi(T-\delta)+Y(e^{-\alpha T}-\beta \Psi)+Z(\Psi\Phi-\lambda), \label{sup2}
\end{equation}
by integrating out the heavy fields, where the coupling $Ye^{-\alpha T}$ can be generated
through non-perturbative effect. The phase monodromy in Eq.~(\ref{phm1}) originates
from the approximate global $U(1)$ symmetry in $W_0$
\begin{equation}
\begin{split}
&\Psi\rightarrow\Psi e^{-iq\te} ~,~\Phi\rightarrow\Phi e^{iq\te} ~,~\\
&Y\rightarrow Y e^{iq\te} ~,~~~T\rightarrow T+i q\te/\alpha~.
\end{split}
\end{equation}
This global $U(1)$ symmetry is exact in the last two terms of $W_0$ while is broken explicitly by its first term,
which is hierarchically smaller but dominates the inflation process.
The phase monodromy of $W_0$ under the circular $U(1)$ rotation is
\begin{equation}
\Psi\rightarrow\Psi e^{-i2\pi},  ~~ W_0\rightarrow W_0+i2\pi\sigma \frac{1}{\alpha}X\Psi.
\end{equation}

The field $X$ during inflation is strongly stabilized at $X=0$ due to the large mass obtained from the factor $e^{X\bar{X}}$
in the $F$-term scalar potential. With this field stabilization the $F$-term scalar potential is simplified as
\begin{equation}
V=e^{\Phi\pb}W_X\bar{W}_{\bar{X}}=a^2e^{r^2}\frac{1}{r^2}((\ln r)^2+\te^2), \label{po1}
\end{equation}
in which $\Phi=r e^{i\theta}$. The potential has interesting helicoid structure, and provides strong stabilization
on the radial component $\langle r\rangle=1$. The phase component of $\Phi$ decouples with the radial component,
 and its Lagrangian after field stabilization becomes
\begin{equation}
L=\partial_\mu\te\partial^\mu \te-ea^2\te^2,
\end{equation}
which gives the quadratic inflation. The inflaton evolves along a helix trajectory--the local valley of helicoid potential.

\subsection{The Natural Inflation}

According to the new Planck results \cite{Ade:2015oja}, natural inflation locates in the region with $95\%$ confidence level
for the effective axion decay constant $f_a\geqslant6.9 ~M_{Pl}$, where $M_{\rm Pl}$ is the reduced Planck scale.
The effective large axion decay constant can be obtained from axion alignment mechanism~\cite{Kim:2004rp}. Nevertheless,
its supergravity or string realization is rather difficult~\cite{Czerny:2014qqa, Kallosh:2014vja}, since generically
the axions (phase) couple with other components in the $F$-term scalar potential, and it is highly non-trivial to
stabilize all the extra components while keep the axions light. The axion alignment with consistent moduli stabilization
was fulfilled in Ref.~\cite{Li:2014lpa}, where the anomalous $U(1)$ gauge symmetry plays a crucial role as
its D-term potential automatically separates the axions from extra components. Inflation based on the anomalous
$U(1)$ gauge symmetry has been studied extensively~\cite{Li:2014owa, Li:2014xna, Mazumdar:2014bna, Wieck:2014xxa,Abe:2014pwa, Domcke:2014zqa}.

We showed that the supergravity setup given by Eq.~(\ref{sup1}) can be slightly modified to accomodate natural
inflation~\cite{Li:2014unh}. With the same K\"ahler potential, we considered the following superpotential
\begin{equation}
W_1=\sigma X\Psi(e^{-\alpha T}-\delta)+Y(e^{-\beta T}-\mu \Psi)+Z(\Psi\Phi-\lambda), \label{sup3}
\end{equation}
in which $1\ll\alpha\ll\beta$ since for each single phase, its decay constant is much lower than the Planck mass,
and a small hierarchy between $\alpha$ and $\beta$ is needed to get super-Planckian phase decay constant.
The last two terms in Eq.~(\ref{sup3}) are the same as these in Eq.~(\ref{sup2}), while the first term,
which is perturbative in Eq.~(\ref{sup2}), now is replaced by the non-perturbative coupling in Eq.~(\ref{sup3}).
And the phase monodromy becomes
\begin{equation}
\Psi\rightarrow\Psi e^{-i2\pi},  ~~ W_1\rightarrow W_1+\sigma X\Psi e^{-\alpha T}(e^{-2\pi i\frac{\alpha}{\beta}}-1).
\end{equation}
By integrating out the heavy fields the superpotential (\ref{sup3}) reduces into
\begin{equation}
W^\prime=a\frac{X}{\Phi}((\Phi^{-b}-c), \label{sup4}
\end{equation}
in which $b=\frac{\alpha}{\beta}\ll1$ and $c\approx1$. Here, the small $b$ arising from the small hierarchy
between $\alpha$ and $\beta$ is crucial to realize large phase decay constant. The potential with fractional power
was introduced in Ref.~\cite{Harigaya:2014eta} to get super-Planckian field excursion and large axion decay constant,
it also plays a key role in Ref.~\cite{Li:2014xna} to obtain the super-Planckian axion decay constant
together with anomalous $U(1)$ gauge symmetry.
And then the scalar potential is
\begin{equation}
\begin{split}
V&=e^{r^2}\frac{a^2}{r^2}(r^{-2b}+c^2-2cr^{-b}\cos(b\te)) \\
&=e^{r^2}\frac{a^2}{r^2}(r^{-b}-c)^2+e^{r^2}\frac{4a^2c}{r^{2+b}}(\sin\frac{b}{2}\te)^2~, \label{po2}
\end{split}
\end{equation}
where the first term has a minimum at $\langle r\rangle=r_0=c^{-\frac{1}{b}}\approx1$ and the minimum of the second term
locates at $\langle r\rangle=r_1=\sqrt{1+\frac{b}{2}}$. Giving $r_0\approx r_1$, the scalar potential has
a global minimum around $r_0\approx1$, where the radial component is well stabilized. The Lagrangian of the phase becomes
\begin{equation}
L=\partial_\mu\te\partial^\mu \te-\Lambda^4\left[1-\cos(b\theta)\right],
\end{equation}
which generates the natural inflation with super-Planckian phase decay constant.

%The helical phase inflation provides simple realizations of both quadratic inflation and natural inflation in supergravity. The phase monodromy corresponds to the explicitly breaking of global $U(1)$ symmetry in the superpotential.  The helix trajectory provides sufficient trans-Planckian field excursion for large field inflation without involving in the physics above the Planck scale, which is likely to introduce dangerous corrections and break the flatness conditions for inflation. However, the quadratic inflation is disfavored by the recent results from Planck observations, predictions on the parameters $n_s$ and $\rr$ from natural inflation locates at the marginal region of Planck results. Therefore the simplest setup for quadratic or natural inflation needs to be modified to agree with the observations.

\section{The Helical Phase Inflation with $U(1)$ symmetry breaking}

The global $U(1)$ symmetry of K\"ahler potential plays a fundamental role to realize the phase inflation in supergravity.
This symmetry provides a flat direction for inflation and resolves the $\eta$ problem. As we pointed out before,
technically the $\eta$ problem can also be solved by an approximate symmetry, {\it i.e.}, the symmetry which protects
the flatness of potential can be explicitly broken, as firstly shown in Ref.~\cite{Li:2013nfa} for the
SUGRA inflation with shift symmetry.

Besides providing a flat direction for inflation, the global $U(1)$ symmetry can protect the phase potential away from
the quantum corrections~\cite{Li:2014unh}. The superpotential is free from quantum loop corrections because of the
non-renormalized theorem, but the correction terms do appear in the K\"ahler potential. In particular,
to generate the effective superpotential like Eqs.~(\ref{sup1}) and (\ref{sup4}), we need to integrate out
the heavy fields above the inflation energy scale. Such process may introduce higher order corrections
to the K\"ahler potential as well. The global $U(1)$ symmetry guarantees that as long as these corrections
do not break $U(1)$ symmetry, they can only slightly affect the field stabilization along the radial direction
instead of modify the phase potential. If the global $U(1)$ symmetry is explicitly broken at tree level,
then the quantum loop effect and heavy fields are likely to generate extra terms that break the $U(1)$ symmetry further,
and then the inflation process may be seriously affected by these corrections depending on the magnitude of the symmetry
breaking term. Fortunately, the slow-roll parameters of inflation
\begin{equation}
\epsilon=\frac{M_P^2}{2}(\frac{V_\phi}{V})^2,~~~\eta=M_P^2\frac{V_{\phi\phi}}{V},
\end{equation}
at the stage when the current universe scale crossed the horizon, are of the order $10^{-2}$.  So the symmetry breaking term
at the order $10^{-2}$ is sufficient to affect the slow-roll parameters. While the higher order corrections from
 the symmetry breaking term are of the order $10^{-4}$ or even smaller, which are far beyond the scope of current observations.

\begin{figure}
\centering
\includegraphics[width=80mm, height=80mm,angle=0]{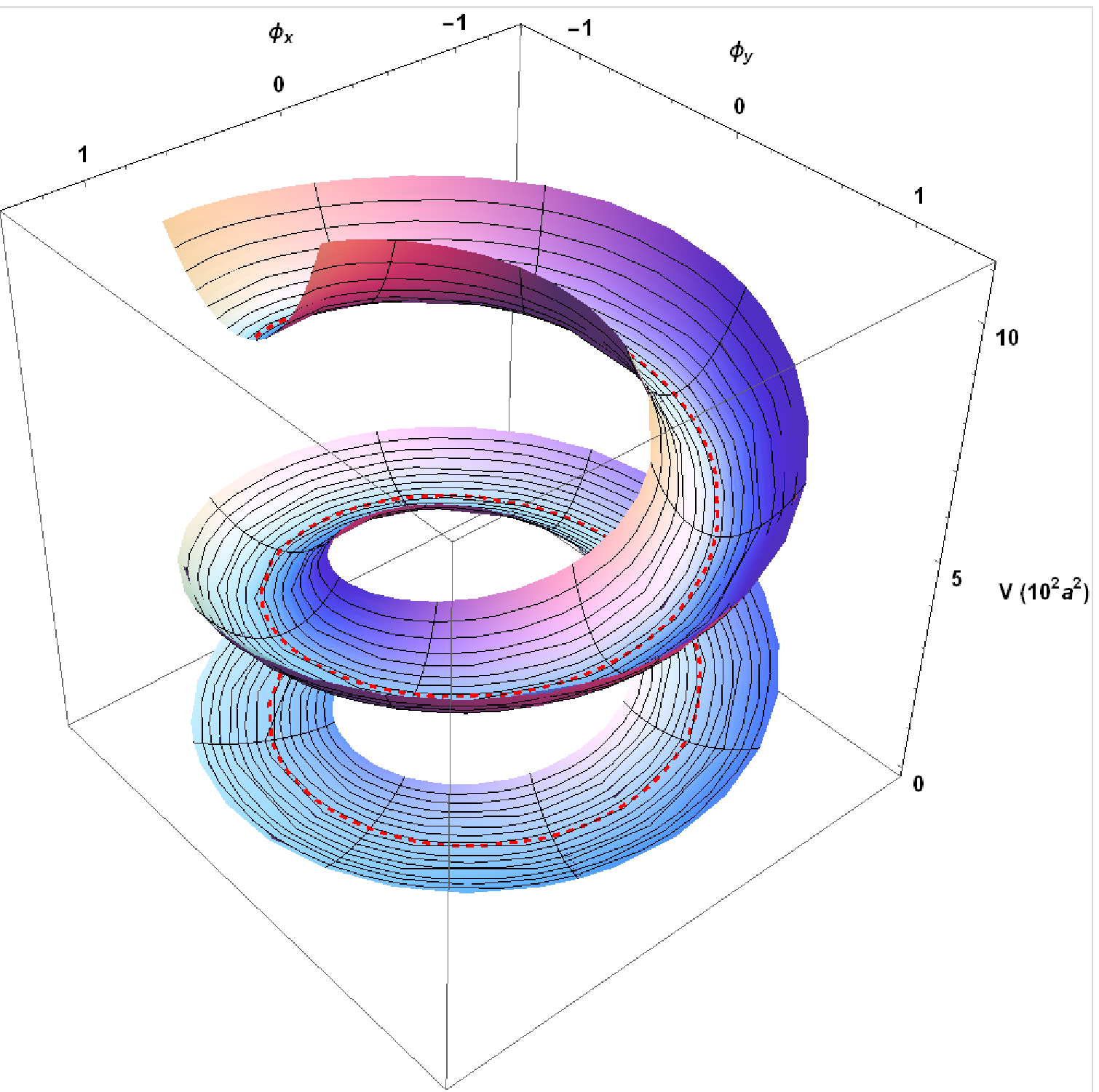}
%\hspace{-.1mm}
\includegraphics[width=80mm, height=80mm,angle=0]{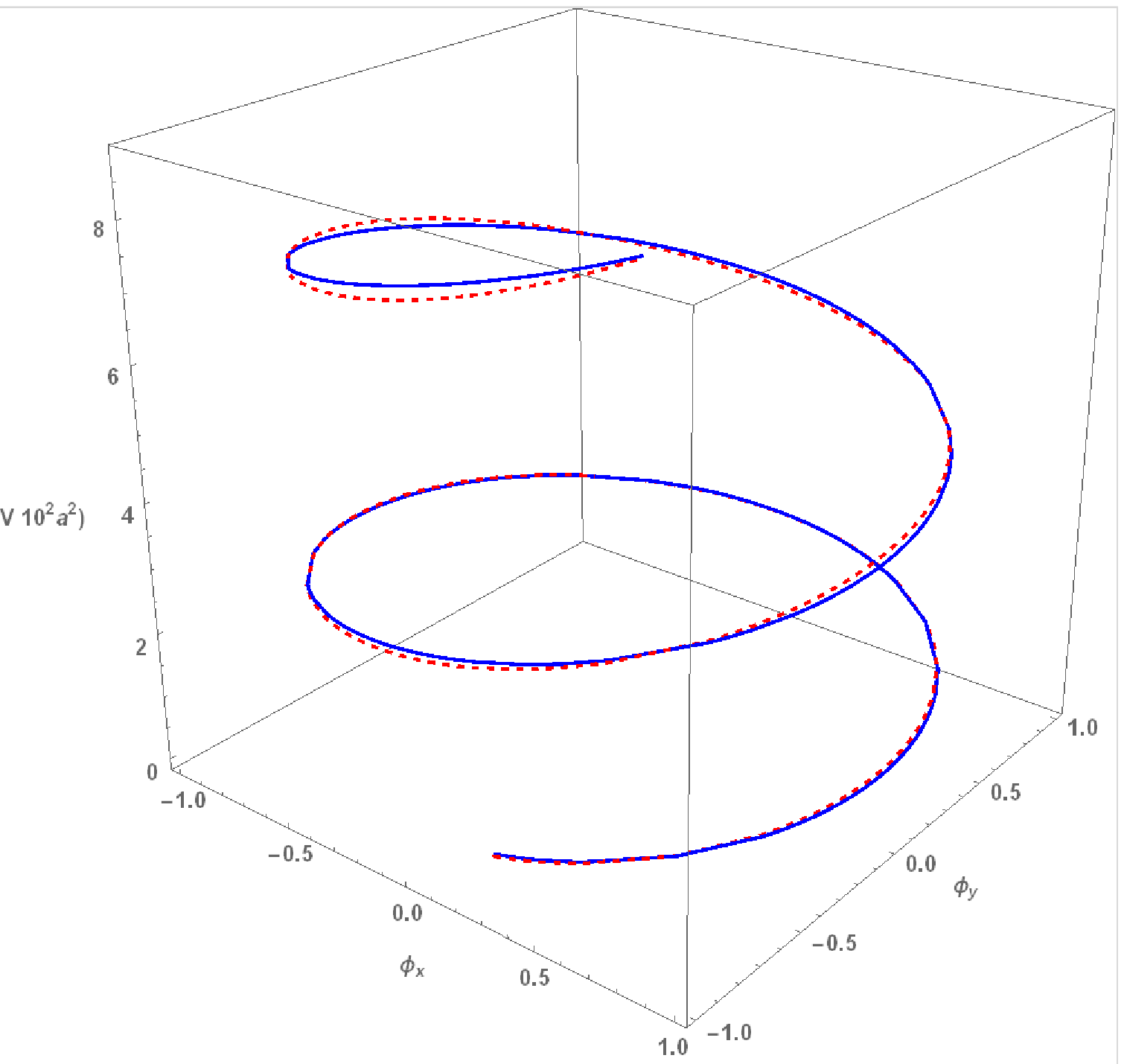}
\caption{Left: the helicoid structure of the potential in Eq.~(\ref{po3}) with $c=0.01$, the red dashed line indicates
the local valley with $r_1=\frac{1}{\sqrt{1+2c\cos(2\theta)}}$, along which the inflaton evolves. Right: the
helix trajectories with broken $U(1)$ symmetry (red dashed) and exact $U(1)$ symmetry (blue).} \label{fig1}
\end{figure}

The global $U(1)$ symmetry can be broken by the real terms like $c(\Phi^n+\pb^n)$ or ${\rm{i}}c(\Phi^n-\pb^n)$.
Taking $c(\Phi^2+\pb^2)$ as an example, we have
\begin{equation}
\begin{split}
K=&\Phi\pb+c(\Phi^2+\pb^2)+X\bar{X}-g(X\bar{X})^2, \\
W=&a\frac{X}{\Phi}\ln\Phi, \label{kah1}
\end{split}
\end{equation}
where the coefficient $c$ is of the order $10^{-2}$. During inflation
$X$ is stabilized at $\langle X\rangle=0$, and then the $F$-term scalar potential is
\begin{equation}
\begin{split}
V&=e^{\Phi\pb+c(\Phi^2+\pb^2)}W_X\bar{W}_{\bar{X}} \\
&=a^2e^{r^2(1+2c \cos(2\theta))}\frac{1}{r^2}((\ln r)^2+\te^2).
\end{split} \label{po3}
\end{equation}
So the global vacuum is $\langle r\rangle=1, \te=0$. For large $\te$ during inflation, the coefficient of above potential
has a minimum at $r_1=\frac{1}{\sqrt{1+2c\cos(2\theta)}}$. Thus, the potential possesses a
helicoid structure, as shown in Fig.~\ref{fig1}.

\begin{figure}
\centering
\includegraphics[width=100mm, height=62mm,angle=0]{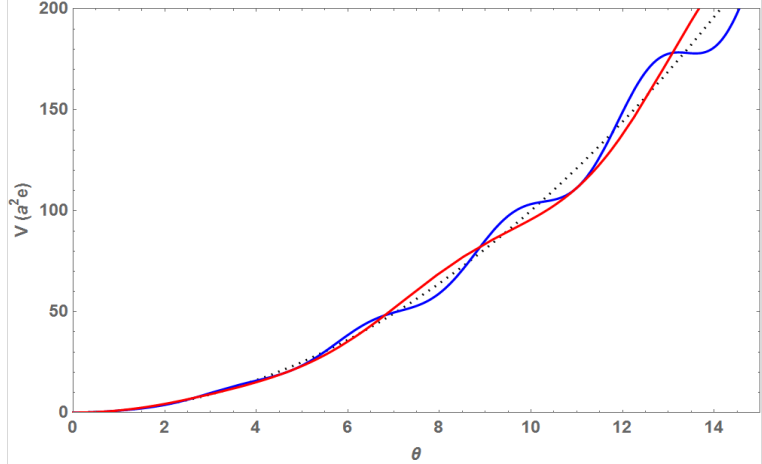}
%\hspace{-.1mm}
\caption{The phase potentials for exact $U(1)$ symmetry (dotted) and explicitly $U(1)$ symmetry breaking (blue and red).
The blue and red curves are respectively the potentials modulated by cosine and sine factors, which originate
from different $U(1)$ symmetry breaking terms. Here, the coefficient of $U(1)$ symmetry breaking term is $c=0.04$. }\label{fig2}
\end{figure}

\begin{figure}
\centering
\includegraphics[width=80mm, height=50mm,angle=0]{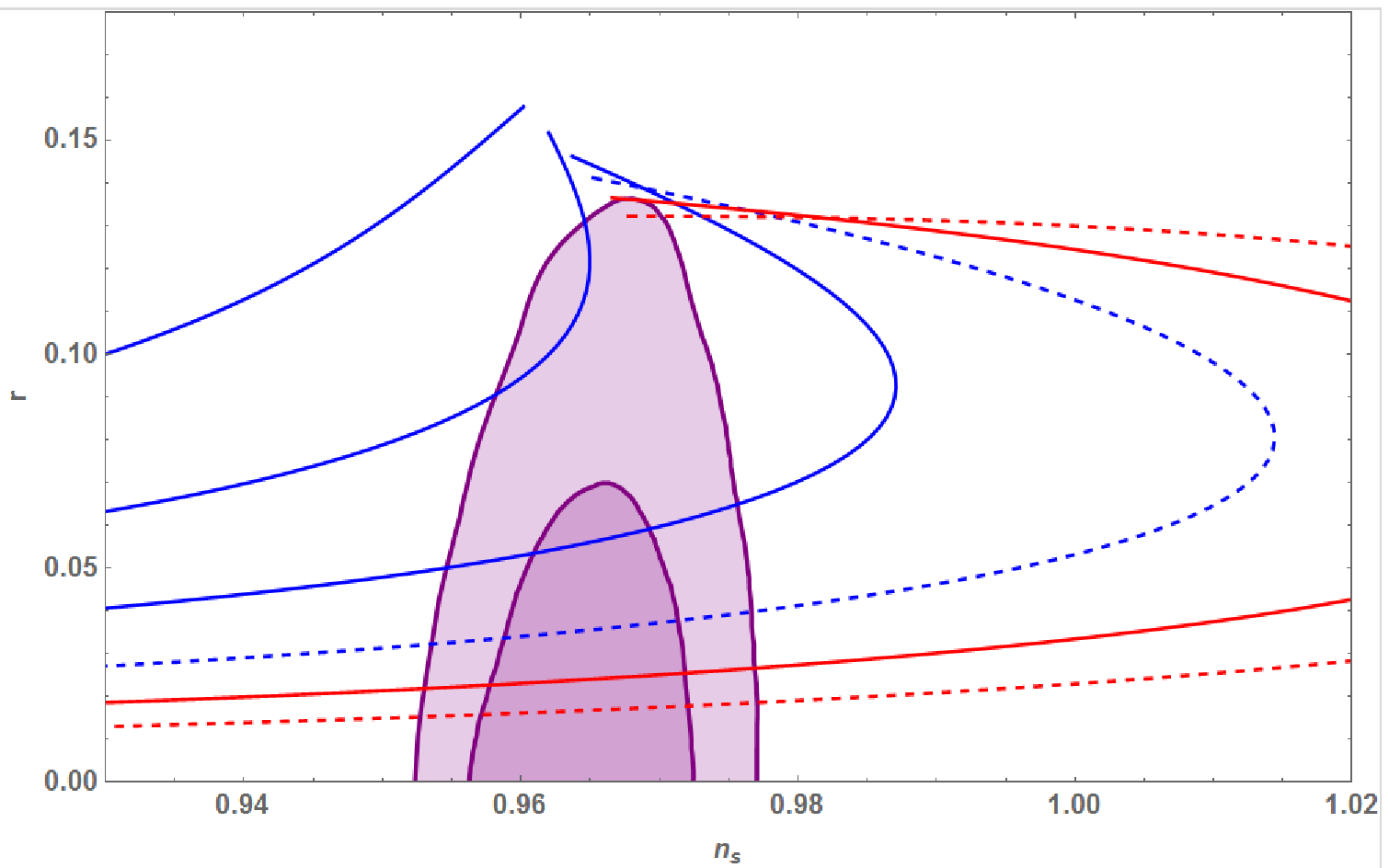}
%\hspace{-.1mm}
\includegraphics[width=80mm, height=51mm,angle=0]{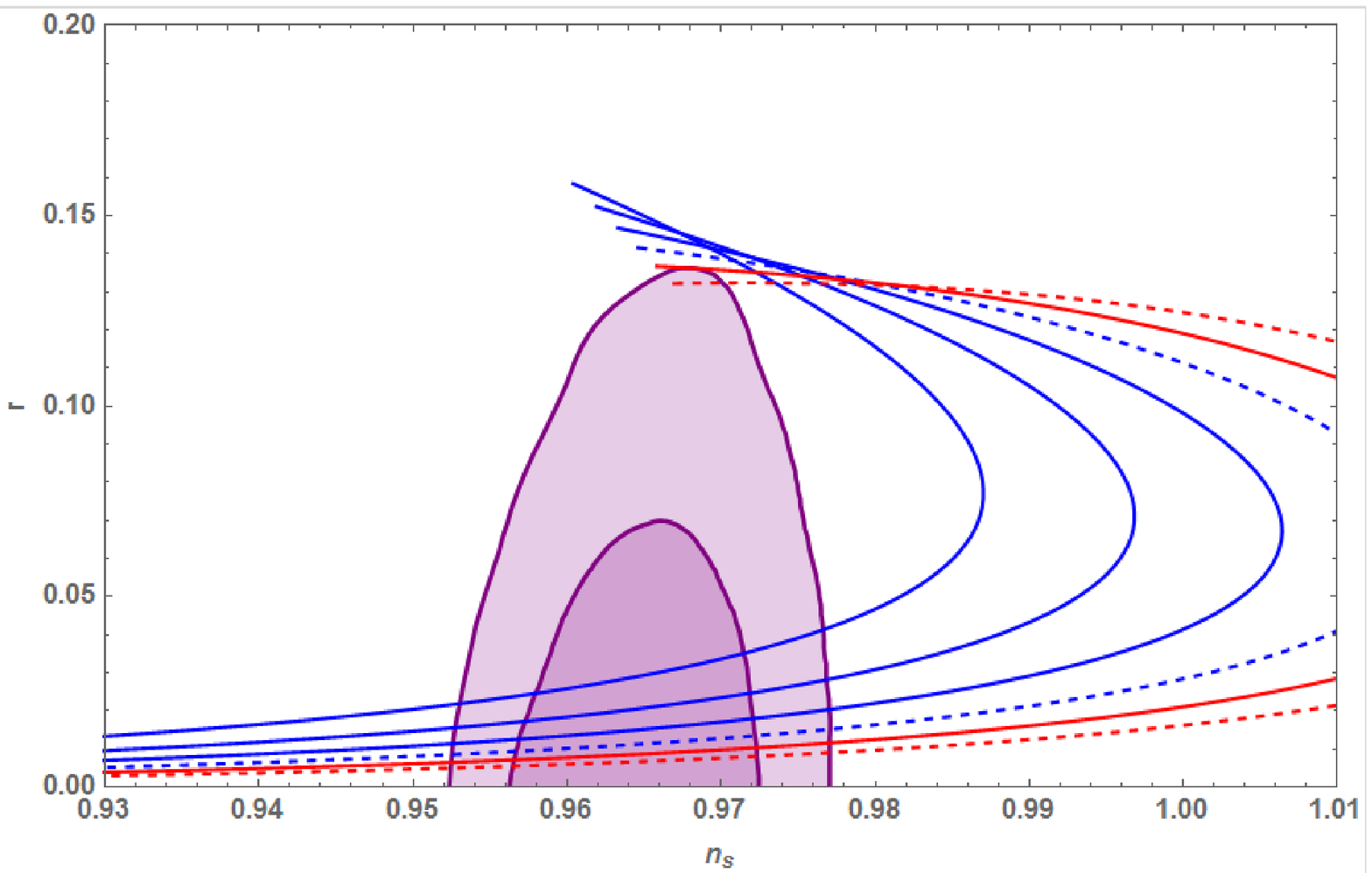}
\caption{$r$ versus $n_s$ for the cosine modulated quadratic inflation with quadratic symmetry breaking term (right)
and the sine modulated quadratic inflation with linear symmetry breaking term (left). The two purple regions
represent the $95\%$ and $68\%$ C.L. according to the Planck results \cite{Ade:2015oja}.
 In each graph, from the left-top to right-bottom, the curves present the $n_s-r$ relations with different
e-folding numbers $N=\{50, 52, 54, 56, 58, 60\}$. The curves start from the predictions of exact quadratic inflation
corresponding to $c=0$. The results in the left graph, which relate to the quadratic $U(1)$ symmetry breaking term,
strongly depend on the e-folding number $N$, indicating the notable scalar spectral index running. In contrast,
for the results from linear $U(1)$ breaking term, as shown in the right graph, they just slightly shift
for different $N$ and correspondingly, the running of spectral index is rather small.}\label{fig3}
\end{figure}

During inflation $r\approx r_1$, the first term $(\ln r)^2\simeq c^2\sim10^{-4}\ll\te^2\sim O(10)$ is negligible. Applying
the radial component stabilization to Eq.~(\ref{po3}), we obtain the phase Lagrangian
\begin{equation}
L=\frac{1}{1+2c \cos(2\theta)}\partial_\mu\theta\partial^\mu\theta-a^2e(1+2c \cos(2\theta))\te^2~. \label{po4}
\end{equation}
 The radial component depends on $\te$ and then is slowly changing during inflation, nevertheless its kinetic energy is
of the order $c^2$ and dropped in above formular.
The quadratic phase potential is modulated by a cosine factor because of the $U(1)$ symmetry breaking term.
The modulated potential is presented in Fig.~\ref{fig2}.  The effects of the potential modulation on inflation
have been widely studied before~\cite{Feng:2003mk, Kobayashi:2010pz,
Takahashi:2013tj, Czerny:2014wua, Wan:2014fra, Higaki:2014sja,Minor:2014xla, Abe:2014xja,Flauger:2014ana, delaFuente:2014aca, Higaki:2015kta}. One of the most interesting effects resulted from the modulation is the running spectral index,
and more results on the running spectral index will be provided later.  In our model, the $U(1)$ symmetry breaking term
also slightly modulates the phase kinetic term.  The predicted inflationary observables can be significantly improved
by the modulation, as presented in Fig.~\ref{fig3}. In particular, the tensor-to-scalar ratio spreads in a broad range.
The $n_s-\rr$ relations strongly depend on the e-folding number, indicating a notable running of spectral index $\alpha_s$.
Moreover, we can  get the small running of spectral index as well.

Next, we shall consider the $U(1)$ symmetry breaking by the linear term ${\rm{i}}c(\Phi-\pb)$.
The K\"ahler potential in Eq.~(\ref{kah1}) is replaced by
\begin{equation}
K=\Phi\pb-{\rm i}c(\Phi-\pb)+X\bar{X}-g(X\bar{X})^2. \label{kah2}
\end{equation}
The scalar potential after field stabilization $X=0$ is
\begin{equation}
V=a^2e^{r^2+2cr\sin\te}\frac{1}{r^2}((\ln r)^2+\te^2).
\end{equation}
The radial component is stabilized at $r\approx1-\frac{c}{2} \sin\te$ up to the order $c$ ($O(c)$),
and then the phase Lagrangian becomes
\begin{equation}
L=\frac{1}{1+c\sin\te}\partial_\mu\theta\partial^\mu\theta-a^2e(1+2c\sin\te)\te^2. \label{po5}
\end{equation}
The corrections from higher order terms $O(c^{n}), n\geqslant2$  on the inflationary observables
are far beyond the current observations so can be ignored. Different from the potential in Eq.~(\ref{po4}),
the above potential is modulated by a sine factor instead of cosine factor as well as a factor 2 difference
on modulation.

The predictions of the inflationary observables from potentials in Eqs.~(\ref{po4}) and (\ref{po5}) are
given in Fig.~\ref{fig3}. Both models can nicely agree with the new Planck observations with $U(1)$ symmetry
breaking parameter $c$ in certain range.
For the potential in Eq.~(\ref{po4}), the results are altered significantly from $N\sim50$ to $N\sim60$.
The potential modulation from quadratic $U(1)$ symmetry breaking term has introduced notable running
of scalar spectral index. In the regions at $68\%$ C.L., for example with $n_s\in[0.96, 0.97]$ and $\rr\simeq0.04$,
the model predicts the running of spectral index $\alpha_s\simeq-0.019$ for $N=56$.
While for the model with linear $U(1)$ symmetry breaking term, the results are much different.
It predicts a rather smaller running of spectral index, which is about $\alpha_s\simeq-0.005$ or even closer to
zero depending on the ranges of $n_s$ and $\rr$. The reason might be that the modulation for potential in Eq.~(\ref{po4})
depends on $\cos2\theta$, while that for potential in Eq.~(\ref{po5}) depends on $\sin\theta$.

\begin{figure}
\centering
\includegraphics[width=80mm, height=65mm,angle=0]{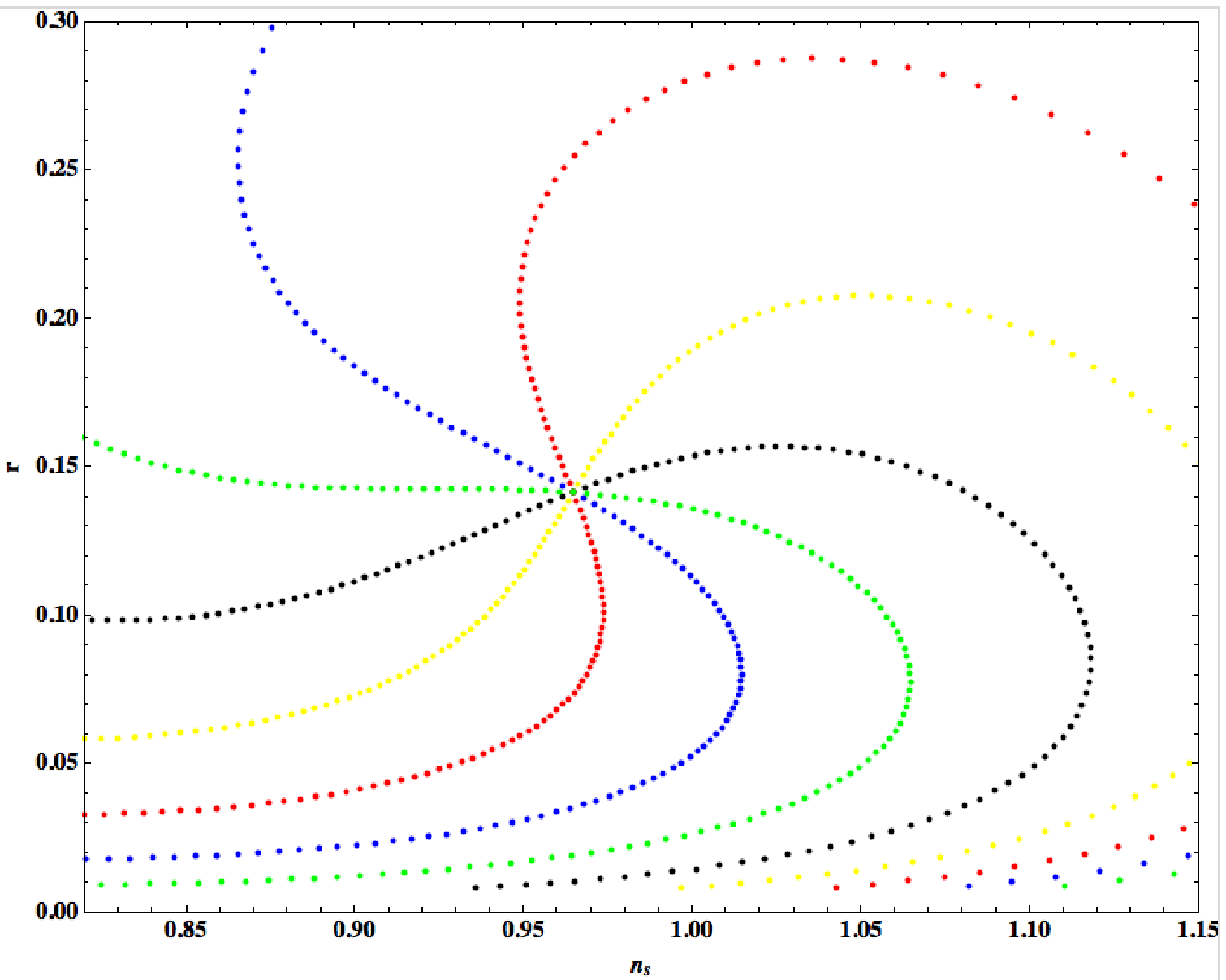}
%\hspace{-.1mm}
\includegraphics[width=80mm, height=64mm,angle=0]{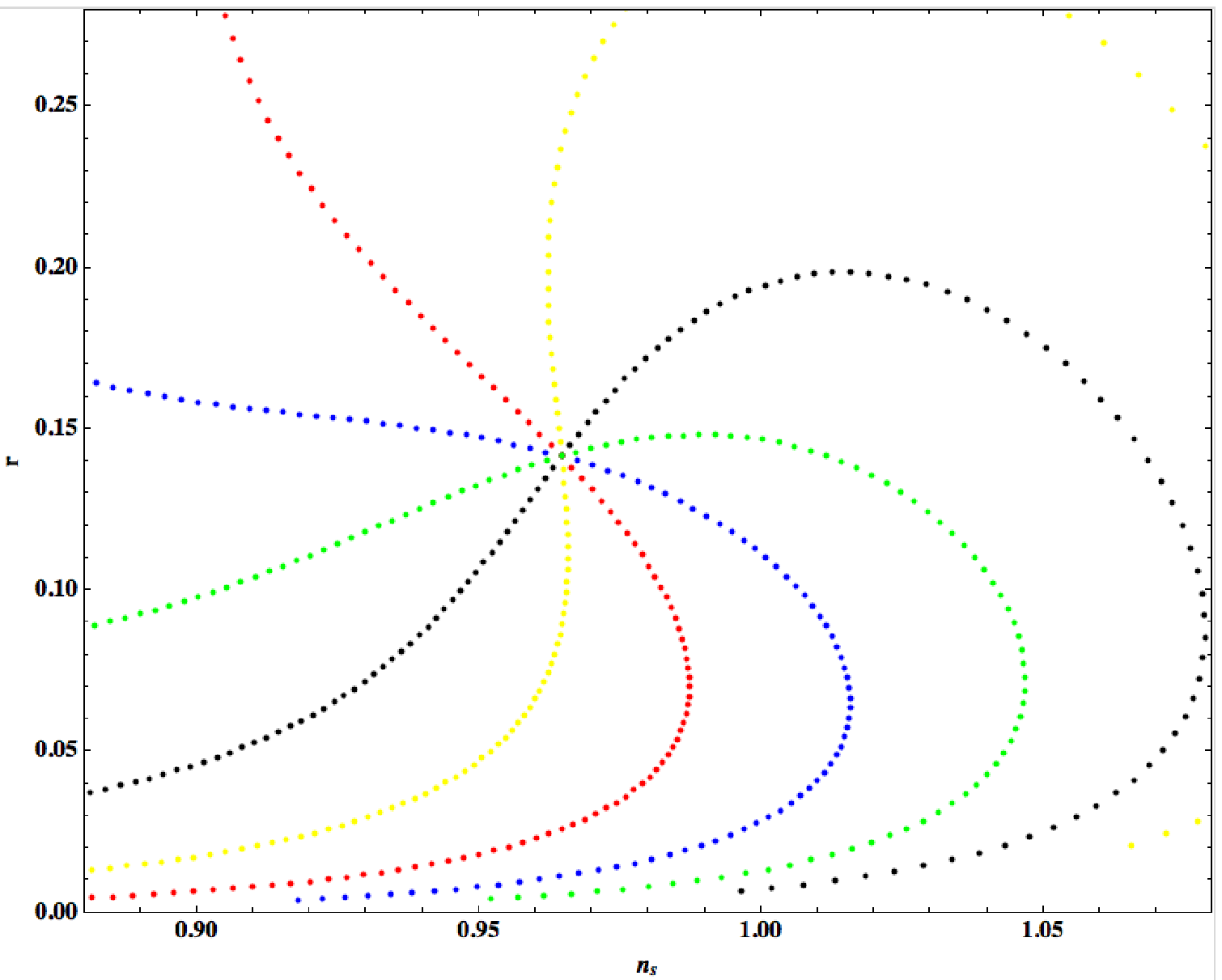}
\caption{$r$ versus $n_s$.
Left: the inflaton potential in Eq.~(\ref{po6}) with $\phi_0=\{0,\frac{\pi}{5},\frac{2\pi}{5},\frac{3\pi}{5},\frac{4\pi}{5}\}$
for the curves with colors of blue, red, yellow, black, green, respectively. The e-folding number is fixed at $N=56$.
Right:  the inflaton potential in Eq.~(\ref{po7}). All the curves intersect at a point of exact quadratic inflation
with the same e-folding number.}\label{fig6}
\end{figure}

The two $U(1)$ symmetry breaking terms $c(\Phi^n+\pb^n)$ and $ic(\Phi^n-\pb^n)$ are related by a discrete phase shift
$\te\rightarrow\te+\frac{\pi}{2n}$. Instead of adopting different $U(1)$ symmetry breaking forms,
such phase shift can also be fulfilled  by introducing an extra phase parameter in the superpotential
\begin{equation}
W=a\frac{X}{\Phi}\ln\frac{\Phi}{\Lambda},
\end{equation}
where $\Lambda=e^{i\phi_0}$. It gives a continuous phase shift and the quadratic potential is modulated by a trigonometric factor.
Starting from the same K\"ahler potentials in Eqs.~(\ref{kah1}) or (\ref{kah2}), with above superpotential
we can finally get the phase Lagrangians up to the order $c$ ($O(c)$)
\begin{equation}
L=\frac{1}{1+2c \cos(2\theta-\phi_0)}\partial_\mu\theta\partial^\mu\theta-a^2e(1+2c \cos(2\theta-\phi_0))\te^2, \label{po6}
\end{equation}
or
\begin{equation}
L=\frac{1}{1+c \sin(\theta-\phi_0)}\partial_\mu\theta\partial^\mu\theta-a^2e(1+2c \sin(\theta-\phi_0))\te^2, \label{po7}
\end{equation}
where $\phi_0$ is a constant and the discrete phase shift $\te\rightarrow\te+\frac{\pi}{2n}$ is included as a special choice
of the constant phase $\phi_0$.
The results of the potentials in Eqs.~(\ref{po6}) and (\ref{po7}) are given in Fig.~\ref{fig6}, where the $n_s-\rr$ curves are
 estimated with a fixed e-folding number $N=56$. For general $N\in[50, 60]$, the curves will be slightly modified
with the shifted intersection point.

\begin{figure}
\centering
\includegraphics[width=80mm, height=50mm,angle=0]{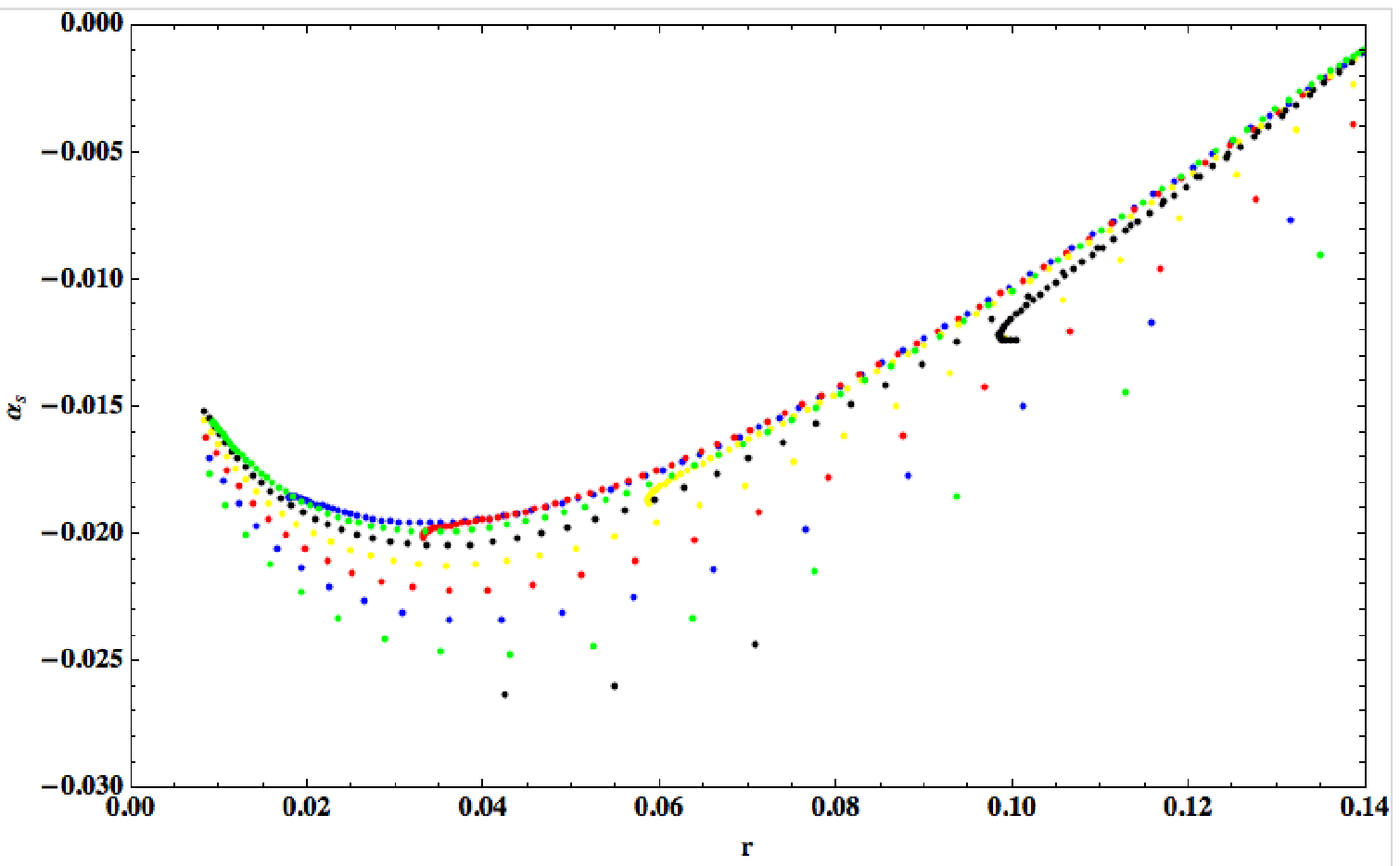}
%\hspace{-.1mm}
\includegraphics[width=80mm, height=50mm,angle=0]{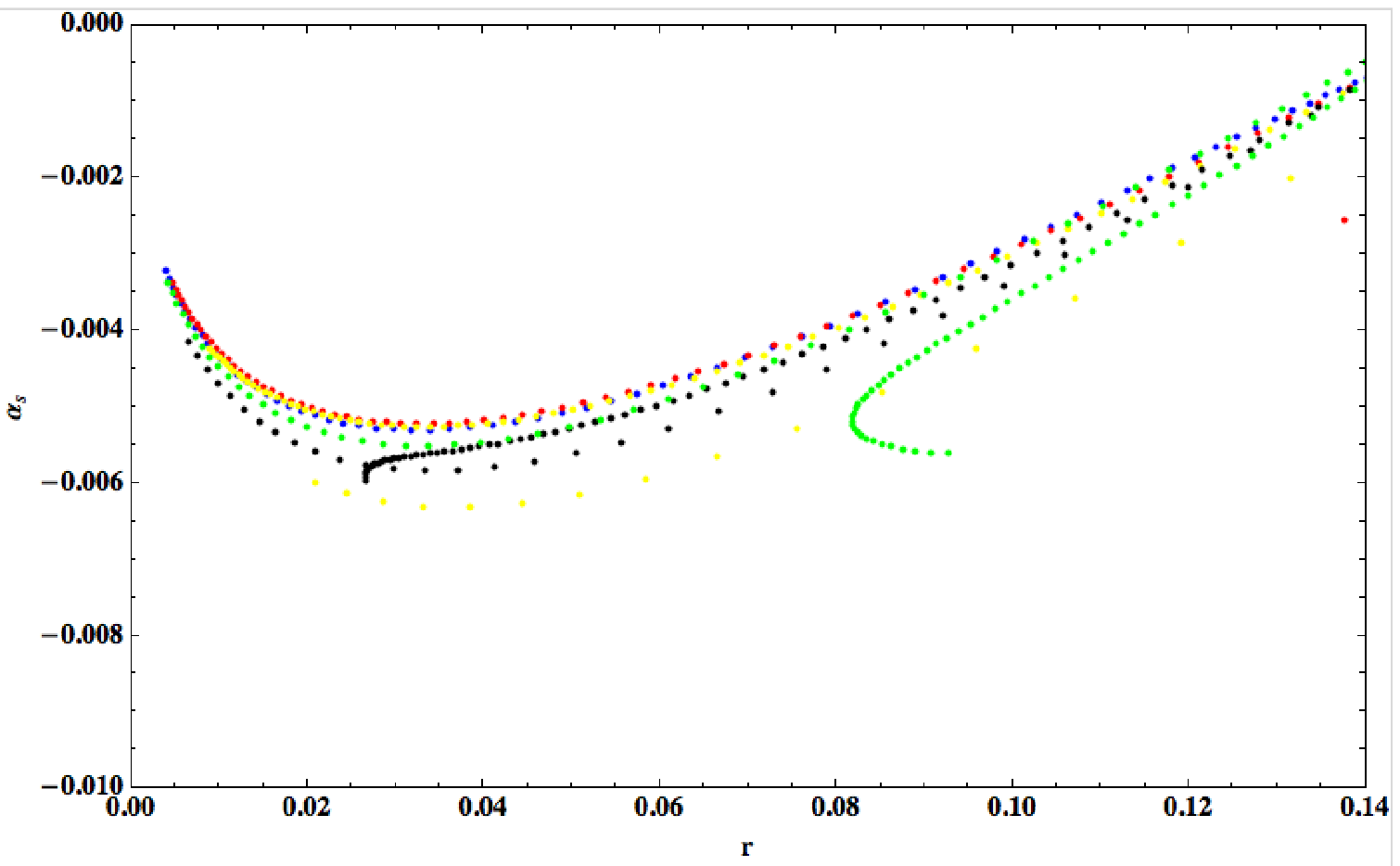}
\caption{The running of spectral index $\alpha_s$ versus $r$ for the inflaton potentials in Eqs.~(\ref{po6}) (left)
and (\ref{po7}) (right) with the fixed e-folding number $N=56$. The color coding is the same
as in Fig.~\ref{fig6}.}\label{fig7}
\end{figure}

The major difference between the potentials in Eqs.~(\ref{po6}) and (\ref{po7}) appears in the running spectral index $\alpha_s$.
As shown in Fig.~\ref{fig7}, for $\rr<0.10$ and fixed $N=56$, the running spectral index $\alpha_s$ generated by
the potential in Eq.~(\ref{po6}) is about $-0.02\sim-0.01$, while the running spectral index is around $-0.006\sim-0.002$
for that in Eq.~(\ref{po7}). So the different $U(1)$ symmetry breaking terms can be clearly distinguished from
the observations of running spectral index. Such a small running spectral index is preferred according to
the Planck observations~\cite{Planck:2015xua, Ade:2015oja}, although a conclusive result is still absent.
Future observations on the running spectral index will determine the $U(1)$ symmetry breaking term of such kind of models.

%\subsection{$U(1)$ Symmetry Breaking for Natural Inflation}
%The $U(1)$ symmetry breaking term can also be introduced in the K\"ahler potential for natural inflation. The supergravity setup is similar to that for quadratic inflation(\ref{quad}) with the superpotential replaced by (\ref{sup4})
%\begin{equation}
%\begin{split}
%K=&\Phi\pb+c(\Phi^2+\pb^2)+X\bar{X}-g(X\bar{X})^2, \\
%W^\prime=&a\frac{X}{\Phi}(\Phi^{-b}-d),
%\end{split}
%\end{equation}
%where $c\sim10^{-2}$, $b\ll1$ and $d\approx1$.
%Through field stabilization $X\rightarrow\langle X\rangle=0$, the F-term scalar potential becomes
%\begin{equation}
%V=e^{r^2(1+2c\cos 2\te)}\frac{a^2}{r^2}(r^{-b}-d)^2+e^{r^2(1+2c\cos 2\te)}\frac{4a^2d}{r^{2+b}}(\sin\frac{b}{2}\te)^2~. \label{po5}
%\end{equation}
%During inflation the second term dominates the inflation potential, which has the local minimum $r=\sqrt{\frac{1+b/2}{1+2c\cos2\te}}$, and the Lagrangian of phase is
%\begin{equation}
%L=\frac{1+b/2}{1+2c \cos(2\theta)}\partial_\mu\theta\partial^\mu\theta-\Lambda^4(1+2c \cos(2\theta))^{1+b/2}(1-\cos b\te).
%\end{equation}
%The $U(1)$ breaking term modulates the scalar potential of natural inflation, which is given in Fig.~\ref{fig2}.  Predictions of the modulated potential with $b=0.1$ is provided in Fig.~\ref{fig4}. As expected, the results can be considered as further ``deformation" of the quadratic inflation, with the $\rm{n_s}-\rr$ curve shifted to the left.

In addition, one can study the $U(1)$ symmetry breaking effect for natural inflation.
The original inflaton potential $V=\Lambda^4(1-\cos b\te)$ is also modulated by a trigonometric factor as below
\begin{equation}
V\propto (1+2c\cos2\te)^{1+\frac{b}{2}}(1-\cos b\te),
\end{equation}
or a similar form depending on the $U(1)$ symmetry breaking term.
The inflationary predictions are similar to the above modulated quadratic inflations, except a slight shift of limitation
related to $c\rightarrow 0$.
Because in such kind of model there are two free parameters (phase decay constant and symmetry breaking coefficient)
while gives the similar results as the modulated quadratic inflation presented in Fig.~\ref{fig3} which only
has one free parameter,
it seems to be less attractive at current stage.

\section{The Shift Symmetry Breaking in Light of Planck 2015}

The shift symmetry as a solution to $\eta$ problem for supergravity inflation was
proposed in \cite{Kawasaki:2000yn}. We first suggested that
by breaking the shift symmetry in the K\"ahler potential, one can get a  broad range
of tensor-to-scalar ratio $\rr$ without changing the scalar spectral index $n_s$~\cite{Li:2013nfa} .
The potential role of the symmetry breaking term was discussed in Ref.~\cite{Kallosh:2010ug}, and
such a symmetry breaking model could generate $\rr$ as large as $\rr\sim0.20$ \cite{Harigaya:2014qza}.

In Ref.~\cite{Li:2014unh}, it was shown that the helical phase inflation can reduce to
the supergravity realization of quadratic inflation with shift symmetry~\cite{Kawasaki:2000yn}.
From the inflationary model given by Eq.~(\ref{sup1}), taking the field redefinition $\Phi=e^{\Psi}$,
the K\"ahler potential and superpotential become
\begin{equation}
\begin{split}
K=&e^{\Psi+\bar{\Psi}}+\cdots=1+\Psi+\bar{\Psi}+\frac{1}{2}(\Psi+\bar{\Psi})^2+\cdots, \\
W=&aX\Psi e^{-\Psi}. \label{sup5}
\end{split}
\end{equation}
The higher order terms have no contribution to the inflation after field stabilization:
$|\Phi|=e^{\rm{Re}(\Psi)}=1,~\rm{Re}(\Psi)=0$. Through a K\"ahler transformation
\begin{equation}
K(\Psi,\bar{\Psi})\rightarrow K(\Psi,\bar{\Psi})+F(\Psi)+\bar{F}(\bar{\Psi}),  ~~~W\rightarrow e^{-F(\Psi)}W,
\end{equation}
the  K\"ahler potential and superpotential in Eq.~(\ref{sup5}) reproduce the well-known model proposed
in Ref.~\cite{Kawasaki:2000yn}
\begin{equation}
K=\frac{1}{2}(\Psi+\bar{\Psi})^2+X\bar{X}+\cdots, ~~W=aX\Psi.
\end{equation}

The simple connection between the helical phase inflation and supergravity model with shift symmetry remains
in the symmetry breaking scenario. One simple choice of the K\"ahler potential with explicitly broken shift symmetry is
\begin{equation}
K={\rm{i}}c(\Psi-\bar{\Psi})+\frac{1}{2}(\Psi+\bar{\Psi})^2+X\bar{X}+\cdots.
\end{equation}
According to the field redefinition $\Phi=e^{\Psi}$, one can easily figure out the
corresponding K\"ahler potential in the supergravity inflation with $U(1)$ symmetry
\begin{equation}
K={\rm i}c(\ln \Phi-\ln\bar{\Phi})+\Phi\bar{\Phi}+X\bar{X}+\cdots.
\end{equation}
The $U(1)$ symmetry breaking term ${\rm i}c(\ln \Phi-\ln\bar{\Phi})$ linearly depends on the phase and introduces the expected factor $e^{-2c\te}$ on the quadratic phase potential. Through a K\"ahler transformation the $U(1)$ symmetry can be restored, the equivalent model is
\begin{equation}
K=\Phi\bar{\Phi}+X\bar{X}+\cdots, ~~~ W=aX\Phi^{b}\ln\Phi,
\end{equation}
where the power $b=-1+{\rm{i}}c$ is complex.
However, the physical origin of the complex power is rather obscure and it is more natural to break
the global $U(1)$ symmetry in the way discussed above.

\subsection*{The Linear Potential in the Shift Symmetry Breaking Scenario}

We shall study the linear potential modulated by an exponential factor, which can be easily realized
in the supergravity inflation with breaking shift symmetry, and comfortably fits with the new Planck results.
Let us start from the following K\"ahler potential and superpotential~\cite{Li:2013nfa}
\begin{equation}
\begin{split}
K&={\rm{i}}\frac{c}{\sqrt{2}}(\Phi-\pb)+\frac{1}{2}(\Phi+\pb)^2+X\bar{X}-g(X\bar{X})^2,  \\
W&=aX\sqrt{\Phi}. \label{sup6}
\end{split}
\end{equation}
Following the usual procedure, the field $X$ is stabilized at $X\rightarrow \langle X\rangle=0$,
and the $F$-term scalar potential becomes
\begin{equation}
V(\sigma,~\chi)=e^K|W_X|^2=e^{-c\sigma+\chi^2}\sqrt{\sigma^2+\chi^2}, \label{lin1}
\end{equation}
in which the complex field $\Phi$ is replaced by $\Phi=\frac{1}{\sqrt{2}}(\sigma+\rm{i}\chi)$.
The imaginary component obtains a mass above the Hubble scale and runs into the global minimum
$\langle\chi\rangle=0$ rapidly. So we get the exponentially modulated linear potential
\begin{equation}
V(\sigma)=e^K|W_X|^2=e^{-c\sigma} |\sigma|. \label{lin2}
\end{equation}
By taking a different superpotential $W=aX\Phi$ in Eq.~(\ref{sup6}), we obtain
 the inflaton potential $V(\sigma)=e^{-c\sigma}\sigma^2$, which is the quadratic potential with exponential modulation.

\begin{figure}
\centering
\includegraphics[width=100mm, height=61.8mm,angle=0]{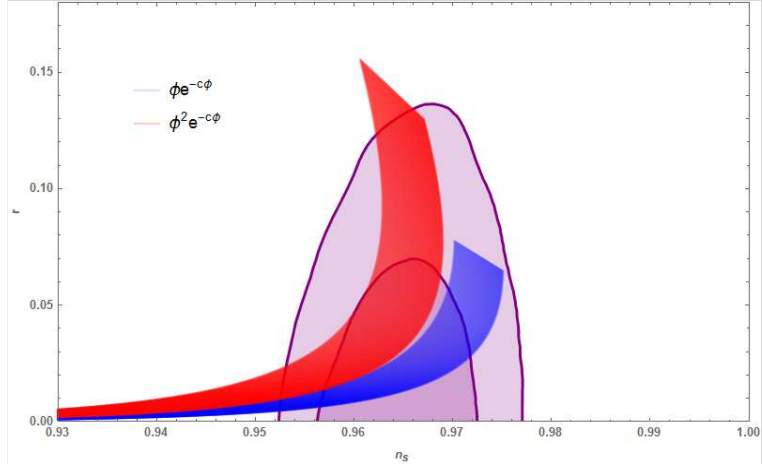}
%\hspace{-.1mm}
\caption{$\rm{n_s}$ versus $\rr$ for the inflationary models with potentials $\phi e^{-c\phi}$ (blue) and
$\phi^2 e^{-c\phi}$ (red). The left-up (right-bottom) boundaries of the strips correspond to the e-folding number
$N=50$ ($N=60$). The purple regions with $68\%$ and $95\%$ C.L. are obtained from the Planck observations.}\label{fig5}
\end{figure}

The predictions of such kind of inflation models are presented in Fig.~\ref{fig5}. Although the quadratic inflation
is disfavoured, especially comparing with the results with B-mode polarizations, its exponential modulated form
remains to be consistent with the new Planck results. The modulated linear potential perfectly agrees
with the observations with or without the B-mode polarizations. With scalar spectral index $n_s\approx0.966$,
it predicts strong tensor fluctuations with tensor-to-scalar ratio $\rr$ around $0.03$, which can be
strictly tested at the future observations.

There are different choices to break the shift symmetry following the idea of Eq.~(\ref{sup6}). For example,
one may introduce a shift symmetry breaking term $c(\Phi-\pb)^2$, together with the same superpotential
the final scalar potential for inflation is $\sigma e^{-c\sigma^2}$ or $\sigma^2 e^{-c\sigma^2}$, which gives
similar $n_s-\rr$ relations as those in Fig.~\ref{fig5}.

\section{Discussions and Conclusion}

We have studied the effects of explicitly symmetry breaking for the supergravity inflation. To solve the $\eta$ problem,
the K\"ahler potential admits an exact global symmetry, either the global $U(1)$ symmetry or shift symmetry,
which can realize the quadratic or natural inflation. However, these simplest setups are disfavoured according
to the new Planck results. We found that by introducing small symmetry breaking term
the inflationary predictions are significantly improved, in the meanwhile the $\eta$ problem remains absent
 even though the symmetry is approximate.

For the supergravity inflation with global $U(1)$ symmetry, the inflaton is the phase component of a complex field
which evolves along helix trajectory during inflation. The global $U(1)$ symmetry plays a crucial role to
protect the inflation dynamics away from the quantum loop corrections. It also realizes the super-Planckian
field excursion without involving in the physics abvoe the Planck scale. Because the $U(1)$ symmetry is broken
explicitly at tree level, the quantum loop corrections will introduce higher order corrections. Fortunately,
to explain the Planck new results, the needed $U(1)$ symmetry breaking term is of the order $10^{-2}$,
and then the higher order corrections are much smaller and ignorable. The $U(1)$ symmetry breaking term
modulates the simple quadratic potential with a cosine or sine factor, consequently, the
$n_s-\rr$ relation is deformed. For the scalar spectral index range $n_s\in[0.96, 0.97]$,
 the tensor-to-scalar ratio can continuously reduce down to $0.015$ in the model with quadratic $U(1)$ breaking term.
Similarly in the model with linear $U(1)$ symmetry breaking term, the tensor-to-scalar ratio falls
into the range $\rr\in[0.006, 0.03]$. The major difference between these two different $U(1)$ symmetry breaking models
is about the running spectral index. The model with quadratic $U(1)$ breaking term predicts a notable running
$\alpha_s\in[-0.02, -0.01]$ for small $\rr<0.08$, in contrast, the model with linear $U(1)$ symmetry breaking term
gives a small running $\alpha_s\in[-0.006, -0.003]$, which comfortably agrees with the new Planck results.
Moreover, we studied the $U(1)$ symmetry breaking model with continuous phase shift. The modulation factor changes
under the phase shift and so are the inflationary observables, the results distribute in the $n_s-\rr$ plane
with a beautiful pattern, and generically they admit deformations with tensor-to-scalar ratio around $\rr\sim0.01$.

In the supergravity inflation with shift symmetry breaking, the inflaton potential is modulated by an exponential factor,
which significantly improves the inflationary predictions as well. Both the quadratic and linear inflation
with shift symmetry breaking can nicely fit with the new Planck results. These models predict
the large tensor fluctuations $\rr\simeq0.04$ with $n_s\simeq0.966$, which can be tested at the near future experiments.

In short, we have shown that the supergravity inflation with broken global symmetry, as a deformation of
the classical quadratic or linear inflation, stays in the region preferred by the new Planck results.
For all these  models there is a remarkable threshold on the tensor fluctuations $\rr\simeq0.01$, which
 is well-known for large field inflation~\cite{Lyth:1996im}. Here it is of special importance
since for the inflationary models obtained from  global symmetry breaking, under the limitation of
scalar spectral index $\rm{n_s}\in[0.96, 0.97]$, the tensor-to-scalar ratio is expected to be of
the order $0.01$. The future experiments may finally tell us whether the Nature adopted
the explicitly symmetry breaking or not.

\begin{acknowledgments}

The work of DVN was supported in part
by the DOE grant DE-FG03-95-ER-40917. The work of TL is supported in part by
    by the Natural Science
Foundation of China under grant numbers 10821504, 11075194, 11135003, 11275246, and 11475238, and by the National
Basic Research Program of China (973 Program) under grant number 2010CB833000.

\end{acknowledgments}


\begin{thebibliography}{99}
%\cite{Guth:1980zm}
\bibitem{Guth:1980zm}
  A.~H.~Guth,
  %``The Inflationary Universe: A Possible Solution to the Horizon and Flatness Problems,''
  Phys.\ Rev.\ D {\bf 23}, 347 (1981);
  %%CITATION = PHRVA,D23,347;%%
  A.~Albrecht and P.~J.~Steinhardt,
  %``Cosmology for Grand Unified Theories with Radiatively Induced Symmetry Breaking,''
  Phys.\ Rev.\ Lett.\  {\bf 48}, 1220 (1982).
  %%CITATION = PRLTA,48,1220;%%

%\cite{Linde:1982}
\bibitem{Linde:1982}
  A.~D.~Linde,
  %``A New Inflationary Universe Scenario: A Possible Solution of the Horizon, Flatness, Homogeneity, Isotropy and Primordial Monopole Problems,''
  Phys.\ Lett.\ B {\bf 108}, 389 (1982);
  
%\cite{Cremmer:1983bf}
\bibitem{Cremmer:1983bf}
  E.~Cremmer, S.~Ferrara, C.~Kounnas and D.~V.~Nanopoulos,
  %``Naturally Vanishing Cosmological Constant In N=1 Supergravity,''
  Phys.\ Lett.\  B {\bf 133}, 61 (1983);
  %%CITATION = PHLTA,B133,61;%%
J.~R.~Ellis, A.~B.~Lahanas, D.~V.~Nanopoulos and K.~Tamvakis,
  %``No-Scale Supersymmetric Standard Model,''
  Phys.\ Lett.\  B {\bf 134}, 429 (1984);
  %%CITATION = PHLTA,B134,429;%%
J.~R.~Ellis, C.~Kounnas and D.~V.~Nanopoulos,
  %``Phenomenological SU(1,1) Supergravity,''
  Nucl.\ Phys.\  B {\bf 241}, 406 (1984);
  %%CITATION = NUPHA,B241,406;%%
%``No Scale Supersymmetric Guts,''
  Nucl.\ Phys.\  B {\bf 247}, 373 (1984);
  %%CITATION = NUPHA,B247,373;%%
A.~B.~Lahanas and D.~V.~Nanopoulos,
  %``The Road to No Scale Supergravity,''
  Phys.\ Rept.\  {\bf 145}, 1 (1987).
  %%CITATION = PRPLC,145,1;%%

%\cite{Li:2014vpa}
\bibitem{Li:2014vpa}
  T.~Li, Z.~Li and D.~V.~Nanopoulos,
  %``Helical Phase Inflation,''
  arXiv:1409.3267 [hep-th].
  %%CITATION = ARXIV:1409.3267;%%
%\cite{Li:2014unh}
\bibitem{Li:2014unh}
  T.~Li, Z.~Li and D.~V.~Nanopoulos,
  %``Helical Phase Inflation and Monodromy in Supergravity Theory,''
  arXiv:1412.5093 [hep-th].
  %%CITATION = ARXIV:1412.5093;%%

%\cite{Kawasaki:2000yn}
\bibitem{Kawasaki:2000yn}
  M.~Kawasaki, M.~Yamaguchi and T.~Yanagida,
  %``Natural chaotic inflation in supergravity,''
  Phys.\ Rev.\ Lett.\  {\bf 85}, 3572 (2000)
  [hep-ph/0004243].
  %%CITATION = HEP-PH/0004243;%%

%\cite{Planck:2015xua}
\bibitem{Planck:2015xua}
  P.~A.~R.~Ade {\it et al.}  [Planck Collaboration],
  %``Planck 2015 results. XIII. Cosmological parameters,''
  arXiv:1502.01589 [astro-ph.CO].
  %%CITATION = ARXIV:1502.01589;%%
%\cite{Ade:2015oja}
\bibitem{Ade:2015oja}
  P.~A.~R.~Ade {\it et al.}  [Planck Collaboration],
  %``Planck 2015. XX. Constraints on inflation,''
  arXiv:1502.02114 [astro-ph.CO].
%\cite{Ade:2015tva}
\bibitem{Ade:2015tva}
  P.~A.~R.~Ade {\it et al.}  [BICEP2 and Planck Collaborations],
  %``A Joint Analysis of BICEP2/Keck Array and Planck Data,''
  %Submitted to: Phys.Rev.Lett.
  [arXiv:1502.00612 [astro-ph.CO]].
  %%CITATION = ARXIV:1502.00612;%%

%\cite{Lyth:1996im}
\bibitem{Lyth:1996im}
  D.~H.~Lyth,
  %``What would we learn by detecting a gravitational wave signal in the cosmic microwave background anisotropy?,''
  Phys.\ Rev.\ Lett.\  {\bf 78}, 1861 (1997)
  [hep-ph/9606387].
  %%CITATION = HEP-PH/9606387;%%






%\cite{Freese:1990rb}
\bibitem{Freese:1990rb}
  K.~Freese, J.~A.~Frieman and A.~V.~Olinto,
  %``Natural inflation with pseudo - Nambu-Goldstone bosons,''
  Phys.\ Rev.\ Lett.\  {\bf 65}, 3233 (1990);
  %%CITATION = PRLTA,65,3233;%%
   F.~C.~Adams, J.~R.~Bond, K.~Freese, J.~A.~Frieman and A.~V.~Olinto,
  %``Natural inflation: Particle physics models, power law spectra for large scale structure, and constraints from COBE,''
  Phys.\ Rev.\ D {\bf 47}, 426 (1993)
  [hep-ph/9207245].
  %%CITATION = HEP-PH/9207245;%%
%\cite{tHooft}
\bibitem{tHooft}
G. ’t Hooft, in {\it Recent Developments in Gauge Theories}, eds. G. ’t Hooft,
{\it et al}, (Plenum Press, New York and London, 1979), p. 135.
%\cite{Li:2013nfa}
\bibitem{Li:2013nfa}
  T.~Li, Z.~Li and D.~V.~Nanopoulos,
  %``Supergravity Inflation with Broken Shift Symmetry and Large Tensor-to-Scalar Ratio,''
  JCAP {\bf 1402}, 028 (2014)
  [arXiv:1311.6770 [hep-ph]].
  %%CITATION = ARXIV:1311.6770;%%

%\cite{Ade:2013uln}
\bibitem{Ade:2013uln}
  P.~A.~R.~Ade {\it et al.}  [Planck Collaboration],
  %``Planck 2013 results. XXII. Constraints on inflation,''
  Astron.\ Astrophys.\  {\bf 571}, A22 (2014)
  [arXiv:1303.5082 [astro-ph.CO]].
  %%CITATION = ARXIV:1303.5082;%%

%\cite{Kallosh:2010ug}
\bibitem{Kallosh:2010ug}
  R.~Kallosh and A.~Linde,
  %``New models of chaotic inflation in supergravity,''
  JCAP {\bf 1011}, 011 (2010)
  [arXiv:1008.3375 [hep-th]].
  %%CITATION = ARXIV:1008.3375;%%

%\cite{Harigaya:2014qza}
\bibitem{Harigaya:2014qza}
  K.~Harigaya and T.~T.~Yanagida,
  %``Discovery of Large Scale Tensor Mode and Chaotic Inflation in Supergravity,''
  Phys.\ Lett.\ B {\bf 734}, 13 (2014)
  [arXiv:1403.4729 [hep-ph]].
  %%CITATION = ARXIV:1403.4729;%%
%\cite{Ade:2014xna}
\bibitem{Ade:2014xna}
  P.~A.~R.~Ade {\it et al.}  [BICEP2 Collaboration],
  %``Detection of B-Mode Polarization at Degree Angular Scales by BICEP2,''
  Phys.\ Rev.\ Lett.\  {\bf 112}, 241101 (2014)
  [arXiv:1403.3985 [astro-ph.CO]].
  %%CITATION = ARXIV:1403.3985;%%

%\cite{German:2001sm}
\bibitem{German:2001sm}
  G.~German, A.~Mazumdar and A.~Perez-Lorenzana,
  %``Angular inflation from supergravity,''
  Mod.\ Phys.\ Lett.\ A {\bf 17}, 1627 (2002)
  [hep-ph/0111371].
  %%CITATION = HEP-PH/0111371;%%

%\cite{Baumann:2010nu}
\bibitem{Baumann:2010nu}
  D.~Baumann and D.~Green,
  %``Inflating with Baryons,''
  JHEP {\bf 1104}, 071 (2011)
  [arXiv:1009.3032 [hep-th]].
  %%CITATION = ARXIV:1009.3032;%%

%\cite{Harigaya:2014eta}
\bibitem{Harigaya:2014eta}
  K.~Harigaya and M.~Ibe,
  %``Simple realization of inflaton potential on a Riemann surface,''
  Phys.\ Lett.\ B {\bf 738}, 301 (2014)
  [arXiv:1404.3511 [hep-ph]].
  %%CITATION = ARXIV:1404.3511;%%

%\cite{McDonald:2014oza}
\bibitem{McDonald:2014oza}
  J.~McDonald,
  %``Sub-Planckian Two-Field Inflation Consistent with the Lyth Bound,''
  JCAP {\bf 1409}, no. 09, 027 (2014)
  [arXiv:1404.4620 [hep-ph]];
  %%CITATION = ARXIV:1404.4620;%%
  J.~McDonald,
  %``A Minimal Sub-Planckian Axion Inflation Model with Large Tensor-to-Scalar Ratio,''
  arXiv:1407.7471 [hep-ph].
  %%CITATION = ARXIV:1407.7471;%%
%\cite{Carone:2014cta}
\bibitem{Carone:2014cta}
  C.~D.~Carone, J.~Erlich, A.~Sensharma and Z.~Wang,
  %``Dante's Waterfall,''
  arXiv:1410.2593 [hep-ph].
  %%CITATION = ARXIV:1410.2593;%%
%\cite{Barenboim:2014vea}
\bibitem{Barenboim:2014vea}
  G.~Barenboim and W.~I.~Park,
  %``Spiral Inflation,''
  arXiv:1412.2724 [hep-ph].
%\cite{McDonald:2014rha, Barenboim:2015zka}
\bibitem{McDonald:2014rha}
  J.~McDonald,
  %``Signatures of Planck Corrections in a Spiralling Axion Inflation Model,''
  arXiv:1412.6943 [hep-ph].
  %%CITATION = ARXIV:1412.6943;%%
%\cite{Barenboim:2015zka}
\bibitem{Barenboim:2015zka}
  G.~Barenboim and W.~I.~Park,
  %``Spiral Inflation with Coleman-Weinberg Potential,''
  arXiv:1501.00484 [hep-ph].




%\cite{Kim:2004rp}
\bibitem{Kim:2004rp}
  J.~E.~Kim, H.~P.~Nilles and M.~Peloso,
  %``Completing natural inflation,''
  JCAP {\bf 0501}, 005 (2005)
  [hep-ph/0409138].
  %%CITATION = HEP-PH/0409138;%%
%\cite{Silverstein:2008sg}
\bibitem{Silverstein:2008sg}
  E.~Silverstein and A.~Westphal,
  %``Monodromy in the CMB: Gravity Waves and String Inflation,''
  Phys.\ Rev.\ D {\bf 78}, 106003 (2008)
  [arXiv:0803.3085 [hep-th]].
  %%CITATION = ARXIV:0803.3085;%%
   L.~McAllister, E.~Silverstein and A.~Westphal,
  %``Gravity Waves and Linear Inflation from Axion Monodromy,''
  Phys.\ Rev.\ D {\bf 82}, 046003 (2010)
  [arXiv:0808.0706 [hep-th]].
  %%CITATION = ARXIV:0808.0706;%%
  L.~McAllister, E.~Silverstein, A.~Westphal and T.~Wrase,
  %``The Powers of Monodromy,''
  JHEP {\bf 1409}, 123 (2014)
  [arXiv:1405.3652 [hep-th]].
  %%CITATION = ARXIV:1405.3652;%%

%\cite{Kaloper:2008fb}
\bibitem{Kaloper:2008fb}
  N.~Kaloper and L.~Sorbo,
  %``A Natural Framework for Chaotic Inflation,''
  Phys.\ Rev.\ Lett.\  {\bf 102}, 121301 (2009)
  [arXiv:0811.1989 [hep-th]].
  %%CITATION = ARXIV:0811.1989;%%
  N.~Kaloper, A.~Lawrence and L.~Sorbo,
  %``An Ignoble Approach to Large Field Inflation,''
  JCAP {\bf 1103}, 023 (2011)
  [arXiv:1101.0026 [hep-th]].
  %%CITATION = ARXIV:1101.0026;%%
  N.~Kaloper and A.~Lawrence,
  %``Natural Chaotic Inflation and UV Sensitivity,''
  Phys.\ Rev.\ D {\bf 90}, 023506 (2014)
  [arXiv:1404.2912 [hep-th]].
  %%CITATION = ARXIV:1404.2912;%%

%\cite{Choi:2014rja}
\bibitem{Choi:2014rja}
  K.~Choi, H.~Kim and S.~Yun,
  %``Natural inflation with multiple sub-Planckian axions,''
  Phys.\ Rev.\ D {\bf 90}, 023545 (2014)
  [arXiv:1404.6209 [hep-th]].
  %%CITATION = ARXIV:1404.6209;%%

%\cite{Tye:2014tja}
\bibitem{Tye:2014tja}
  S.-H.~H.~Tye and S.~S.~C.~Wong,
  %``Helical Inflation and Cosmic Strings,''
  arXiv:1404.6988 [astro-ph.CO].
  %%CITATION = ARXIV:1404.6988;%%

%\cite{Kappl:2014lra}
\bibitem{Kappl:2014lra}
  R.~Kappl, S.~Krippendorf and H.~P.~Nilles,
  %``Aligned Natural Inflation: Monodromies of two Axions,''
  Phys.\ Lett.\ B {\bf 737}, 124 (2014)
  [arXiv:1404.7127 [hep-th]].
  %%CITATION = ARXIV:1404.7127;%%



%\cite{Czerny:2014qqa}
\bibitem{Czerny:2014qqa}
  M.~Czerny, T.~Higaki and F.~Takahashi,
  %``Multi-Natural Inflation in Supergravity and BICEP2,''
  Phys.\ Lett.\ B {\bf 734}, 167 (2014)
  [arXiv:1403.5883 [hep-ph]].
  %%CITATION = ARXIV:1403.5883;%%
%\cite{Kallosh:2014vja}
\bibitem{Kallosh:2014vja}
  R.~Kallosh, A.~Linde and B.~Vercnocke,
  %``Natural Inflation in Supergravity and Beyond,''
  Phys.\ Rev.\ D {\bf 90}, 041303 (2014)
  [arXiv:1404.6244 [hep-th]].
  %%CITATION = ARXIV:1404.6244;%%



%\cite{Li:2014lpa}
\bibitem{Li:2014lpa}
  T.~Li, Z.~Li and D.~V.~Nanopoulos,
  %``Aligned Natural Inflation and Moduli Stabilization from Anomalous $U(1)$ Gauge Symmetries,''
  JHEP {\bf 1411}, 012 (2014)
  [arXiv:1407.1819 [hep-th]].
  %%CITATION = ARXIV:1407.1819;%%

%\cite{Li:2014owa}
\bibitem{Li:2014owa}
  T.~Li, Z.~Li and D.~V.~Nanopoulos,
  %``Chaotic Inflation in No-Scale Supergravity with String Inspired Moduli Stabilization,''
  Eur.\ Phys.\ J.\ C {\bf 75}, no. 2, 55 (2015)
  [arXiv:1405.0197 [hep-th]].
  %%CITATION = ARXIV:1405.0197;%%
%\cite{Li:2014xna}
\bibitem{Li:2014xna}
  T.~Li, Z.~Li and D.~V.~Nanopoulos,
  %``Natural Inflation with Natural Trans-Planckian Axion Decay Constant from Anomalous $U(1)_X$,''
  JHEP {\bf 1407}, 052 (2014)
  [arXiv:1405.1804 [hep-th]].
  %%CITATION = ARXIV:1405.1804;%%
%\cite{Mazumdar:2014bna}
\bibitem{Mazumdar:2014bna}
  A.~Mazumdar, T.~Noumi and M.~Yamaguchi,
  %``Dynamical breaking of shift-symmetry in supergravity-based inflation,''
  Phys.\ Rev.\ D {\bf 90}, no. 4, 043519 (2014)
  [arXiv:1405.3959 [hep-th]].
  %%CITATION = ARXIV:1405.3959;%%
%\cite{Wieck:2014xxa}
\bibitem{Wieck:2014xxa}
  C.~Wieck and M.~W.~Winkler,
  %``Inflation with Fayet-Iliopoulos Terms,''
  Phys.\ Rev.\ D {\bf 90}, no. 10, 103507 (2014)
  [arXiv:1408.2826 [hep-th]].
  %%CITATION = ARXIV:1408.2826;%%
%\cite{Abe:2014pwa}
\bibitem{Abe:2014pwa}
  H.~Abe, T.~Kobayashi and H.~Otsuka,
  %``Towards natural inflation from weakly coupled heterotic string theory,''
  arXiv:1409.8436 [hep-th].
  %%CITATION = ARXIV:1409.8436;%%
%\cite{Domcke:2014zqa}
\bibitem{Domcke:2014zqa}
  V.~Domcke, K.~Schmitz and T.~T.~Yanagida,
  %``Dynamical D-Terms in Supergravity,''
  Nucl.\ Phys.\ B {\bf 891}, 230 (2015)
  [arXiv:1410.4641 [hep-th]].
  %%CITATION = ARXIV:1410.4641;%%

%\cite{Feng:2003mk}
\bibitem{Feng:2003mk}
  B.~Feng, M.~z.~Li, R.~J.~Zhang and X.~m.~Zhang,
  %``An inflation model with large variations in spectral index,''
  Phys.\ Rev.\ D {\bf 68}, 103511 (2003)
  [astro-ph/0302479].
  %%CITATION = ASTRO-PH/0302479;%%

%\cite{Kobayashi:2010pz}
\bibitem{Kobayashi:2010pz}
  T.~Kobayashi and F.~Takahashi,
  %``Running Spectral Index from Inflation with Modulations,''
  JCAP {\bf 1101}, 026 (2011)
  [arXiv:1011.3988 [astro-ph.CO]].
  %%CITATION = ARXIV:1011.3988;%%

%\cite{Takahashi:2013tj}
\bibitem{Takahashi:2013tj}
  F.~Takahashi,
  %``The Spectral Index and its Running in Axionic Curvaton,''
  JCAP {\bf 1306}, 013 (2013)
  [arXiv:1301.2834, arXiv:1301.2834 [astro-ph.CO]].
  %%CITATION = ARXIV:1301.2834,;%%
%\cite{Czerny:2014wua}
\bibitem{Czerny:2014wua}
  M.~Czerny, T.~Kobayashi and F.~Takahashi,
  %``Running Spectral Index from Large-field Inflation with Modulations Revisited,''
  Phys.\ Lett.\ B {\bf 735}, 176 (2014)
  [arXiv:1403.4589 [astro-ph.CO]].
  %%CITATION = ARXIV:1403.4589;%%
%\cite{Wan:2014fra}
\bibitem{Wan:2014fra}
  Y.~Wan, S.~Li, M.~Li, T.~Qiu, Y.~Cai and X.~Zhang,
  %``Single field inflation with modulated potential in light of the Planck and BICEP2,''
  Phys.\ Rev.\ D {\bf 90}, no. 2, 023537 (2014)
  [arXiv:1405.2784 [astro-ph.CO]].
  %%CITATION = ARXIV:1405.2784;%%

%\cite{Higaki:2014sja}
\bibitem{Higaki:2014sja}
  T.~Higaki, T.~Kobayashi, O.~Seto and Y.~Yamaguchi,
  %``Axion monodromy inflation with multi-natural modulations,''
  JCAP {\bf 1410}, no. 10, 025 (2014)
  [arXiv:1405.0775 [hep-ph]].
  %%CITATION = ARXIV:1405.0775;%%

%\cite{Minor:2014xla}
\bibitem{Minor:2014xla}
  Q.~E.~Minor and M.~Kaplinghat,
  %``Inflation that runs naturally: Gravitational waves and suppression of power at large and small scales,''
  arXiv:1411.0689 [astro-ph.CO].
  %%CITATION = ARXIV:1411.0689;%%
%\cite{Abe:2014xja}
\bibitem{Abe:2014xja}
  H.~Abe, T.~Kobayashi and H.~Otsuka,
  %``Natural inflation with and without modulations in type IIB string theory,''
  arXiv:1411.4768 [hep-th].
  %%CITATION = ARXIV:1411.4768;%%
%\cite{Flauger:2014ana}
\bibitem{Flauger:2014ana}
  R.~Flauger, L.~McAllister, E.~Silverstein and A.~Westphal,
  %``Drifting Oscillations in Axion Monodromy,''
  arXiv:1412.1814 [hep-th].
  %%CITATION = ARXIV:1412.1814;%%
%\cite{delaFuente:2014aca}
\bibitem{delaFuente:2014aca}
  A.~de la Fuente, P.~Saraswat and R.~Sundrum,
  %``Natural Inflation and Quantum Gravity,''
  arXiv:1412.3457 [hep-th].
  %%CITATION = ARXIV:1412.3457;%%
%\cite{Higaki:2015kta}
\bibitem{Higaki:2015kta}
  T.~Higaki and F.~Takahashi,
  %``Elliptic Inflation: Interpolating from natural inflation to $R^2$-inflation,''
  arXiv:1501.02354 [hep-ph].
  %%CITATION = ARXIV:1501.02354;%%
\end{thebibliography}
\end{document}